\documentclass[10pt,english]{article}
\usepackage{graphicx}
\usepackage{pstcol,times,color,graphicx,graphics,pstricks,pst-node,pst-coil,pst-grad}
\newpsobject{showgrid}{psgrid}{subgriddiv=1,griddots=10,gridlabels=6pt}
\usepackage{amsmath}
\begin{document}
\title{The motion of two masses coupled to a massive spring.}
\author{F C Santos\footnote{e-mail: filadelf@if.ufrj.br}, Y A Coutinho\footnote{e-mail: yara@if.ufrj.br},
L Ribeiro-Pinto\footnote{e-mail: leandrorp@if.ufrj.br.}, and A C Tort\footnote{e-mail: tort@if.ufrj.br.}\\
Instituto de F\'{\i}sica
\\
Universidade Federal do Rio de Janeiro\\
Caixa Postal 68.528; CEP 21941-972 Rio de Janeiro, Brazil}
\maketitle
\begin{abstract}
We discuss the classical motion of a spring of
arbitrary mass coupled to two arbitrary massive blocks attached at its ends. A
general approach to the problem is presented and some general results are obtained.  Examples for which a simple elastic function can be inferred are discussed and the
normal modes and normal frequencies obtained. An approximation procedure to the evaluation of the normel frequencies in the case of uniform elastic function and mass density is also discussed. 
 \vfill
\end{abstract}
\section{Introduction}
The motion of one or two masses linked by a massless spring  constrained to
move on a straight line and without friction is analysed in several
introductory and undergraduate mechanics textbooks, see for example
\cite{French1971}, \cite{Marion&Thornton1995}. In the case of two arbitrary masses, the two-body problem is solved by a reduction to the problem of a single body oscillating with an angular frequency equal
to $\sqrt{K_e/\mu },$ where $K_e$ is the spring constant and
$\mu $ is the reduced mass of the system, and the motion of the
centre of mass of the system the velocity of which is constant if no external
forces are present. Moreover, since only the masses make contributions to the
kinetic energy and to the total linear momentum, the mechanical energy conservation
theorem and the linear momentum conservation theorem can be applied
without much ado. The forces acting
on the masses are due to the spring deformation at the extremities
where the masses are attached to. This is the reason why Newton's
third law of motion cannot be directly applied to them. We are
forced to consider in more detail the mechanism of interaction between the two masses and in particular their
interaction with the extremities of the spring to which each one of them is
attached to. However, due to the fact that the spring is massless we
can state also that at any given moment of time the sum of those
forces is zero. Then, in an equivalent way, we can think that
the masses move under the action of the force that one mass exerts
on the other, thereby complying with the third law in such a way
that we can ignore the existence of the spring.

The correction to the frequency for the case where one of end of the
spring is kept fixed and the mass $m$ of the
spring, though not zero, is much less than the mass $M$ attached to
the oscillating end is well known. In this case in order to get the angular frequency up to first order we can consider the spring massless and replace the mass of the oscillating
body by an effective mass that is equal to $M+m/3$, see for example \cite{French1971}, see also \cite{GalloniKohen} and references therein.

In this paper we will consider a more general situation. We will
consider the problem of two arbitrary masses, say $M_1$ and $M_2$, attached to a spring of arbitrary mass $m$. The effects caused by the undulatory propagation of the massive spring
deformation along the spring length will be taken into account. Solutions
to particular situations such as the ones described above will be
considered as appropriate limits of a less particular solution. We believe
that the approach we take here may be of some pedagogical value for
advanced students and instructors as well.
\section{The equations of motion for the massive spring and attached masses}
\begin{figure}[!h]
\begin{center}
\begin{pspicture}(-5,-2)(5,3)
%
%
\psset{arrowsize=0.2 2}

\dotnode(-2.0,1.5){A}

\dotnode(2.0,1.5){B}

\psset{coilarm=.01, coilwidth=.3}

\nccoil[linewidth=0.3mm]{A}{B}

\dotnode(-3.0,0.0){A}

\dotnode(4.0,0.0){B}

\psset{coilarm=.01, coilwidth=.3}

\nccoil[linewidth=0.3mm]{A}{B}

\rput(-2.0,2.25){$P(0)$}

\rput(2.0,2.25){$P(\ell)$}

\rput(0.5,2.25){$P(x)$}

\rput(0.0,0.65){$P(x)$}

\psdot(0.5,1.5)




\psline[linewidth=0.2mm]{|->}(-4,-0.65)(0,-.65)

\rput(-2,-1.15){$u\left(x,t\right)$}

\rput(-4,-0.35){$\mathcal{O}$}

\psdot(0.0,0.0)

\end{pspicture}

\caption{The motion of a point of the spring with respect to an
inertial frame is described by the coordinate $u\left(x,t\right)$.
Given a point $P$ of the spring we associated with it the parameter
$x$. This association is independent of the dynamical state of the
spring.}

\label{slinky1}

\end{center}

\end{figure}
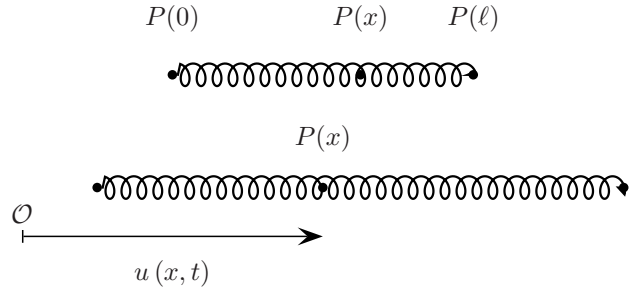

We begin by establishing the equation of motion for the massive
spring along a single spatial dimension. In order to do so we
introduce an auxiliary parameter $x$ that will help us to describe
the properties of the spring such as for example its tension
or its density at a given point. With this end in mind let us consider the
spring in a non-deformed condition and denote by $\ell$ its natural
length. Now we define a one-to-one correspondence between the spring
viewed as a one-dimensional smooth matter distribution and the closed
interval $\left[0,\ell\right]$ in
such a way that $x=0$ corresponds to the left end of the spring and
$x=\ell$ to its right end, see Figure \ref{slinky1}. To an arbitrary point $P$ on the spring there corresponds a point $x\in\left[0,\ell\right]$. The parameter $x$
must not be viewed as a regular spatial coordinate. This parameter
can be thought of if we wish as an internal degree of freedom of the
spring and it is not subject to the transformations associated with
the one-dimensional Galileo group, for instance, non-relativistic boosts or
translations. Were the string made up of $N$ discrete masses labeled
by a discrete index $j$ running from $1$ to $N$, this index would
play a role analogous to $x$. We assume that the correspondence
established here holds for any state of motion of the spring,
exactly as in the case of the discrete model. Now let it be an
inertial reference frame $\mathcal{S}$ and a suitable coordinate
system and let us suppose that the spring moves along the $u$-axis
such that the position of a point of the spring with respect to
$\mathcal{S}$ is given by the function $u\left(x,t\right)$, Figure \ref{slinky1}. The tension ${\rm T}$ at a point of the spring is given by \cite{Landau}
\begin{equation}
{\rm T}\left( x,t\right) =\kappa \left( x\right) \left( \frac{\partial u\left( x,t \right)} {\partial x} -1\right)   \label{leideforca}
\end{equation}
where $\kappa \left( x\right)$ is the elastic function of the
spring which on physical grounds we suppose to be always positive, that is $\kappa (x) > 0$ for any $x\in\left[0,\ell\right]$. In this way, at a given point the force that the right portion of the spring exerts on the left portion wiil be ${\rm T}\left(x,t\right)$ and conversely the force that the left portion of the spring exerts on the right portion will be $-{\rm T}\left(x,t\right)$. Consider now an element of the spring determined by $x$ e
$x+dx$. The resultant force acting on this element is
\begin{eqnarray}
dF \left(x,t\right) &=&{\rm T}\left( x+dx, t\right) -{\rm T}\left( x,t\right) \nonumber \\
&=&\frac{\partial }{\partial x}\left[ \kappa \left( x\right) \left( \frac{%
\partial u\left( x,t\right)} {\partial x}-1\right) \right] dx
\end{eqnarray}
If $\rho\left(x\right)$ is the linear mass density of the spring then upon
applying Newton's second law of motion to the element of mass
$\rho\left(x\right)dx$ we obtain
\begin{equation}
\frac{\partial }{\partial x}\left[ \kappa \left( x\right) \left( \frac{%
\partial u\left( x,t\right) }{\partial x}-1\right) \right] =\rho \left( x\right) \frac{%
\partial ^{2}u\left( x,t\right) }{\partial t^{2}}.  \label{eqondau}
\end{equation}
Equation (\ref{eqondau}) controls the motion of the spring. It can
be simplified by introducing the variable
\begin{equation}
\xi \left( x,t\right) =u\left( x,t\right) -x,  \label{defksi}
\end{equation}
then the equation of motion of the spring becomes
\begin{equation}
\frac{\partial }{\partial x}\left[ \kappa \left( x\right) \frac{\partial \xi
\left( x,t\right) }{\partial x}\right] =\rho \left( x\right) \frac{\partial ^{2}\xi \left( x,t\right) }{\partial
t^{2}}.  \label{eqondaksi}
\end{equation}
A word of caution: though $\xi\left(x,t\right)$ is related to the
deformation of spring it does note represent this deformation
directly.

Notice that the equation of motion of the massive spring in the form given by
equations (\ref{eqondau}) or (\ref {eqondaksi}) is invariant under the
galilean transformations. In fact, if we go from the inertial system
$\mathcal{S}$ to the inertial system $\mathcal{S}^{\,\prime }$ that
moves with velocity $V$ with respect to $\mathcal{S}$ the following
evident relations hold
\begin{eqnarray}
u^\prime\left( x,t\right) &=&u\left( x,t\right) +Vt \\
\frac{\partial u^\prime \left( x,t\right)}{\partial x} &=&\frac{\partial u\left( x,t\right) }{\partial x} \\
\frac{\partial ^{2}u^\prime \left( x,t\right)}{\partial t^{2}} &=&\frac{\partial ^{2}u\left( x,t\right) }{%
\partial t^{2}}.
\end{eqnarray}
In this sense the galilean invariance of the equation of motion of the
spring is manifest in accordance with the fact that this equation derives from a
straightforward application of the principles of newtonian
mechanics.

Let us now consider the coupled masses. Let us model them by means
of two point particles one of mass $M_{1}$ coupled to the left end
$(x=0)$ of the spring and the other of mass $M_{2}$ coupled to the right end
($x=\ell$). Making use of equation (\ref{eqondaksi}) and Newton's
second and third laws we can write the equations of motion of the
masses as
\begin{eqnarray}
M_{1}\frac{\partial ^{2}\xi \left( 0,t\right)}{\partial t^{2}} &=&\kappa
\left( 0\right) \frac{\partial \xi \left( 0,t\right)}{\partial x}
\label{cc1ksi} \\
M_{2}\frac{\partial ^{2}\xi \left(\ell,t\right)}{\partial t^{2}}
&=&-\kappa \left(\ell\right) \frac{\partial \xi \left(
\ell,t\right)}{\partial x} \label{cc2ksi}
\end{eqnarray}
The complete solution of these equations and of equation
(\ref{eqondaksi}) demands that we prescribe the initial conditions
\begin{eqnarray}
\xi \left( x, t=0\right) &=&\varphi \left( x\right) - x \label{cini1} \\
\frac{\partial \xi \left( x,t=0\right)}{\partial t} &=&\psi \left( x\right)
\label{cini2}
\end{eqnarray}
where $\varphi\left(x\right):=u\left(x,0\right)$ and $\psi\left(x\right):=\partial u\left(x,t=0\right)/\partial t$ describe the initial position and velocity of the points of the spring. 

Our aim is to obtain a general solution $\xi\left(x,t\right)$ -- or $ u\left(x,t\right) $ -- to the problem and therefore describe an arbitrary state of motion of the system i.e., the two blocks plus the massive string. 

\section{General solution of the equations of motion}
We begin by solving equation (\ref{eqondaksi}) by the method of separation of variables, that is, we look for a solution of the form
\begin{equation}
\xi \left( x,t\right) =X\left( x\right) T\left( t\right)
\label{solxt}
\end{equation}
that satisfy also the boundary conditions given by equations (\ref{cc1ksi}) and (\ref{cc2ksi}).
Taking equation (\ref{solxt}) into   (\ref{eqondaksi}) and
introducing the separation constant $-\lambda$ we have
\begin{eqnarray}
\frac{d^{2}T(t)}{dt^{2}}+\lambda T(t) &=&0  \label{eqt} \\
\frac{d}{dx}\left[\kappa (x)\frac{dX(x)}{dx}\right] +\lambda \rho
(x) X(x) &=&0\label{eqx}
\end{eqnarray}
Equation (\ref{cc1ksi}) imposes a boundary condition on equation
(\ref{eqx}) to obtain it we take equation  (\ref{solxt}) into equation (\ref{cc1ksi}) and write
\begin{equation}
M_{1}X\left( 0\right) \frac{d^{2}T(t)}{dt^{2}}=\kappa\left( 0\right)
T(t)\frac{dX\left( 0\right) }{dx}
\end{equation}
and taking into account equation (\ref{eqt}) we obtain
\begin{equation}
\kappa\left( 0\right) X^{\prime }\left( 0\right) =-\lambda
M_{1}X\left( 0\right) \label{cc1}
\end{equation}
In the same way, taking equation  (\ref{solxt}) into (\ref{cc2ksi}) and combining with equation (\ref{eqt}) we obtain the
condition
\begin{equation}
\kappa\left(\ell\right) X^{\prime }\left(\ell\right) =\lambda
M_{2}X\left( \ell\right) \label{cc2}
\end{equation}
Let us show now that the eigenvalue $\lambda$ cannot assume negative values. Suppose that we have an eigenfunction $X(x)$ corresponding to a particular eigenvalue $\lambda$. Consider the following identity which can be derived after an integration by parts and use of equations (\ref{cc1}), (\ref{cc2}), and (\ref{eqx})
\begin{equation}
\int_0^\ell\,\kappa\left(x\right)X^{\,\prime 2}(x)\,dx=\lambda\left[
M_1X^2\left(0\right)+M_2X^2\left(\ell\right))+\int_o^\ell\,\rho\left(x\right) X^2\left(x\right)\,dx\right]
\end{equation}
Since the left-hand side is always non-negative and the bracket on the right-hand side is always positive we conclude that $\lambda$ is non-negative. 

The null eigenvalue is physically acceptable and has a special meaning. The reason is that $\lambda=0$ is common to all springs regardless of their mass density, elastic function and the masses of the attached particles at their extremities. Notice that the corresponding eigenfunction (the zero mode) can be obtained in a general form. This particular mode is not associated with an oscilatory motion of the spring.  In fact, for $\lambda=0$ 
the temporal function has the form
\begin{equation}
T(t)=\alpha_0 t+\beta_0 ,
\end{equation}
where $\alpha_0 $ and $\beta_0 $ are constants. On the other hand,
equation (\ref{eqx}) for $\lambda=0$ yields
\begin{equation}
\frac{dX(x)}{dx}=\frac{\gamma}{\kappa\left( x\right) }
\end{equation}
where $\gamma$ is an integration constant.  Boundary conditions as expressed by equations (\ref{cc1ksi}) and (\ref{cc2ksi}), or equivalently equations
(\ref{cc1}) and (\ref{cc2}), demand $\gamma=0$, so that $X\left(
x\right) =b=\mbox{constant}$. The eigenfunction corresponding to this
eigenvalue is then
\begin{eqnarray}
\xi_0\left( x,t\right) &=& b \left(\alpha_0 t+\beta_0 \right)\nonumber
\\ &=& x_{0}+V t
\end{eqnarray}
where we have introduced the new constants $x_0$ and $V$. It is clear that this solution corresponds to a uniform motion of
the entire system (masses plus spring) with a common velocity $V$.
The zero mode motion is related to galilean boosts and may be added to any other solution of the problem if questions about galilean invariance are an issue.

Finally, let us consider the case of positive $\lambda$. Setting $\lambda=\omega^2$ for convenience we write the solutions to equation (\ref{eqt}) as
\begin{equation}\label{Tn}
T_n\left(t\right)=A_n\cos\left(\omega_n t + \phi_n\right)
\end{equation}
where $n$ is a positive integer, $\omega_n$ is the $n$-th frequency eigenvalue indexed in crescent order ($\omega_1 < \omega_2 < \omega_3 \dots$) and $A_n$ and $\phi_n$ are constants. The $n$-th eigensolution to equation {{\ref{eqondaksi}) corresponding to the $n$-th eigenfrequency is
\begin{equation}
\xi_n\left(x,t\right)=X_n\left(x\right) T_n\left(t\right)
\end{equation}
These modes represent the oscillatory modes of the system.
The general solution can be written as
\begin{equation}\label{generalxi}
\xi\left( x,t\right) =x_{0} + Vt + \sum_{n=1}^\infty\,X_n\left( x\right)T_n\left(t\right)
\end{equation}
Consequently, in terms of the function $u\left(x,t\right)$ the general solution will be given by
\begin{equation}\label{generalu}
u\left(x,t\right)=x_0+x+Vt+\sum_{n=1}^\infty\,X_n\left( x\right)T_n\left(t\right)
\end{equation}

The next step is the explicit determination of the spectrum of eigenfrequencies
$\omega$. This is, however, a hard task to perform and in principle it can be
accomplished only if we also know explicitly the elastic function
$\kappa\left(x\right)$. As mentioned before, the zero mode is the only mode that does not depend on the form of $\kappa \left(x\right)$. 
\section{The orthogonality of the eigenfunctions}
Before dealing with concrete examples let us consider a little bit more some of the formal aspects of our problem. Equations (\ref{cc1}) and (\ref{cc2}) can be read as boundary
conditions for equation (\ref{eqx}) therefore only
for certain values of $\lambda$ there will be solutions to this
equation. The reader will recognise immediately that we are dealing with a
Sturm-Liouville system. Let us consider then two different
eigenvalues, say $\lambda _{m}$ and $\lambda _{n}$ and their
respective eigenfunctions $X_{m}\left( x\right)$ and $X_{n}\left(
x\right)$. These eigenfunctions satisfy the differential equations
\begin{eqnarray}
\frac{d}{dx}\left[\kappa\left(x\right)\frac{dX_{m}\left(x\right)}{dx}\right]
+\lambda _{m}\rho X_{m}\left(x\right) &=& 0
\label{eqxm} \\
\frac{d}{dx}\left[\kappa\left(x\right)\frac{dX_{n}\left(x\right)}{dx}\right]
+\lambda _{n}\rho X_{n}\left(x\right) &=& 0 \label{eqxn}
\end{eqnarray}
As usual we multiply the first equation by  $X_{n}$ and the second
by $X_{m}$, subtract one from the other and after simple additional
manipulations we end up with
\begin{equation}
\left(\lambda _{m}-\lambda _{n}\right) \rho\left(x\right)
X_{m}\left(x\right)X_{n}\left(x\right)+\frac{d}{dx}\left[\kappa\left(x\right)\left(
X_{n}\left(x\right)\frac{dX_{m}\left(x\right)}{dx}-X_{m}\left(x\right)\frac{dX_{n}\left(x\right)}{dx}\right)
\right] =0 \label{eqmn}
\end{equation}
Integrating this last equation over the domain $\left[0
,\ell\right]$ and taking into account the boundary conditions given
by equations  (\ref{cc1}) and (\ref{cc2}) we obtain after some
simplifications
\begin{equation}
\left( \lambda _{m}-\lambda _{n}\right) \left[
\int_{0}^{l}\rho\left(x\right)
X_{m}\left(x\right)X_{n}\left(x\right) dx+ \text{ }M_{2}X_{m}\left(
l\right) X_{n}\left( l\right) +M_{1}X_{m}\left( 0\right)
X_{n}(0)\right] =0.
\end{equation}
At this point we define a scalar product in the space of functions
that  will be convenient for our purposes. Let the functions $f\left(
x\right) $ and $g\left( x\right) $ be defined in the closed interval $\left[0
,l\right] $, then by definition their scalar product is
\begin{equation}
\left\langle f ,g \right\rangle
=\int_{0}^{l}\rho (x) f\left( x\right) g\left( x\right) dx+\text{ }M_{2}f\left(
l\right) g\left( l\right) +M_{1}f\left( 0\right) g(0).  \label{esc}
\end{equation}
With this definition for the scalar product we can consider the
eigenfunctions corresponding to different eigenvalues as an 
orthonormal set of eigenfunctions, i.e.
\begin{equation}
\langle X_m, X_n \rangle=\delta_{mn}, \;\;\;\;  m,n=0,1,2,\dots
\end{equation}
By making use of the initial conditions and the above orthonormality condition the determination of the constants $A_n$ and $\phi_n$ that appear in equation (\ref{Tn}) and therefore in the general solution can be done in a systematic way. For the zero mode, for instance, we have
\begin{equation}
X_0(x)=\frac{1}{\sqrt{M_1+M_2+M_{\mbox{\tiny spring}} }}
\end{equation}

\section{Conservation laws}
Linear momentum and mechanical energy conservation theorems can be proven under quite general conditions. The former depends on the fact that the system is isolated and the latter depends also on the fact that the internal forces can be considered as conservatives.
Let us consider first the linear momentum of the system. Our goal
will be to determine explicitly the contribution of the massive
spring to the total linear momentum.

The linear momentum due to the two blocks is given by
\begin{equation}
P_{\mbox{\tiny blocks}}=M_{1}\frac{\partial u\left( 0,t\right)}{\partial t} +M_{2}\frac{\partial u\left( \ell,t\right)%
}{\partial t}   \label{qmovp}
\end{equation}
Making use of equations (\ref{cc1ksi}) and (\ref{cc2ksi}), we can
recast the total time derivative of $P_{\mbox{\tiny blocks}}$ into
the form
\begin{equation}
\frac{d\,P_{\mbox{\tiny blocks}} }{dt}=k\left(0\right)
\frac{\partial \xi\left(0,t\right)}{\partial x} -k\left(\ell\right)
\frac{\partial \xi\left( \ell,t\right)}{\partial x} \label{mmassas}
\end{equation}
On the other hand we can integrate equation (\ref{eqondaksi}) over
the domain $\left[ 0,\ell\right] $ to obtain
\begin{equation}
\frac{d^{2}}{dt^{2}}\int_{0}^{\ell}\rho \left( x\right) \xi\left(
x,t\right) dx=k\left( \ell\right) \frac{\partial \xi\left( \ell,t\right)}{\partial
x} -k\left( 0\right) \frac{\partial \xi\left( 0,t\right)}{\partial
x} \label{monda1}
\end{equation}
Taking this result into equation (\ref{mmassas}) we have
\begin{equation}
\frac{d}{dt}\left( P_{\mbox{\tiny blocks}}+\int_{0}^{l}\rho \left(
x\right) \frac{\partial \xi\left( x,t\right) }{\partial t}dx\right) =0.
\end{equation}
Defining the linear momentum of the spring by
\begin{equation}
P_{\mbox{\tiny spring}}=\int_{0}^{l}\rho \left( x\right) \frac{\partial u\left( x,t\right) }{\partial t}=\int_{0}^{l}\rho \left( x\right) \frac{\partial \xi\left( x,t\right) }{\partial t}
dx  \label{monda3}
\end{equation}
we see that the total linear momentum of the system $P_{\mbox{\tiny
blocks}}+P_{\mbox{\tiny spring}}$ is conserved. The total linear
momentum can be rewritten in the form
\begin{equation}
P_{\mbox{\tiny total}}=
M_{1}\frac{d u_{1}\left(t\right)}{dt}+M_{2}\frac{d u_{2}\left(t\right)}{dt}+\int_{0}^{l}\rho
\left( x\right) \frac{\partial u\left( x,t\right)}{\partial t} dx
\end{equation}
where $u_{1}\left(t\right)\equiv u\left(0,t\right)$ and
$u_{2}\left(t\right)\equiv u\left(\ell,t\right)$. In terms of $\xi\left(x,t\right)$ we have
\begin{equation}
P_{\mbox{\tiny total}}=
M_{1}\frac{d \xi_{1}\left(t\right)}{dt}+M_{2}\frac{d \xi_{2}\left( t\right)}{dt}+\int_{0}^{l}\rho
\left( x\right) \frac{\partial \xi\left( x,t\right)}{\partial t} dx
\end{equation}
From equation (\ref{generalxi}) or (\ref{generalu}) we can rewrite the total linear momentum in the form
\begin{equation}
P_{\mbox{\tiny total}}=\left(M_1+M_2+m\right)V+\sum_{n=1}^\infty\left[M_1X_n\left(0\right)+M_2X_n\left(\ell\right)+\int_0^\ell\,\rho\left(x\right)X_n\left(x\right)\,dx\right]\dot T\left(t\right)
\end{equation}
This expression can be rewritten in the form
\begin{equation}
P_{\mbox{\tiny total}}=\left( M_1+M_2+m \right)V+\sum_{n=1}^\infty\,\langle X_m, X_0\left(x\right) \rangle \dot T(\left(t\right)
\end{equation}
Since $X_0$ and $X_n$ are orthogonal we see that only the zero mode contributes to the total linear momentum
\begin{equation}
P_{\mbox{\tiny total}}=\left( M_1+M_2+M_{\mbox{\tiny spring}} \right)V
\end{equation}
From this result we see that the constant $V$ is the velocity of the centre of mass of the system, as expected.

We now consider the mechanical energy of the system. The kinetic
energy of the blocks is given by
\begin{equation}
T_{1}+T_{2}=\frac{1}{2}M_{1}\left(\frac{\partial u\left(
0,t\right)}{\partial t}\right)^2 +\frac{1}{2}M_{2}\left(\frac{\partial u\left(
\ell,t\right)}{\partial t}\right)^2  \label{ecinm1m2}
\end{equation}
which evidently is not \emph{per se} a conserved quantity because the
blocks exchange energy with the spring. It follows that in order to
have conservation of the mechanical energy it is mandatory that any
variation of the kinetic energy of the blocks be compensated by a
variation of the  energy of the spring, kinetic, potential or both.
Keeping this in mind we derive (\ref{ecinm1m2}) with respect to the
time to obtain
\begin{equation}
\frac{d}{dt}\left( T_{1}+T_{2}\right) =M_{1}\frac{\partial u\left( 0,t\right) }{\partial t}%
\frac{\partial ^{2}u\left( 0,t\right)}{\partial t^{2}}
+M_{2}\frac{\partial u\left(\ell,t\right)}{\partial t} \frac{\partial ^{2}u\left(\ell,t\right)}{%
\partial t^{2}}
\end{equation}
Combining this result with equations (\ref{cc1ksi}) and (\ref{cc2ksi})
we can eliminate the masses of the blocks and write
\begin{equation}
\frac{d}{dt}\left( T_{1}+T_{2}\right) =\kappa\left( 0\right) \frac{\partial u\left( 0,t\right)}{%
\partial t} \left[ \frac{\partial u\left(0
,t\right)}{\partial x} -1\right] -\kappa\left(\ell\right)\frac{\partial
u\left(\ell,t\right)}{\partial t} \left[ \frac{\partial u\left(\ell,t\right)}{\partial
x} -1\right] \label{energia1}
\end{equation}
We can recast this equation into a more useful form if we first multiply equation (\ref{eqondau}) by $\partial u\left( x,t\right)/\partial t $ to obtain
\begin{equation}
\rho \left( x\right) \frac{\partial u\left( x,t\right)}{\partial t}\frac{\partial ^{2}u\left( x,t\right)}{%
\partial t^{2}}=\frac{\partial }{\partial x}\left[\kappa\left(x\right)\left(\frac{\partial
u\left( x,t\right)}{\partial x}-1\right) \frac{\partial u\left( x,t\right)}{\partial t}\right]
-\frac{1}{2}\kappa\left(x\right) \frac{\partial }{\partial t}\left(
\frac{\partial u\left( x,t\right)}{\partial x}-1\right) ^{2}
\end{equation}
Then integrating this result over the interval $\left[ 0,\ell\right] $ we
will have
\begin{eqnarray}
\kappa\left(\ell\right) \left(\frac{\partial u\left(\ell,t\right)}{\partial x} -1%
\right) \frac{\partial u\left(\ell,t\right)}{\partial t}
-\kappa\left(0\right)
\frac{\partial u\left( 0,t\right) }{\partial x}\frac{\partial u\left( 0,t\right) }{\partial t}%
&=&\frac{d}{dt}\int_{0}^{l}\frac{1}{2}\rho \left(
x\right) \left( \frac{\partial u\left( x,t\right)}{\partial t}\right) ^{2}dx
\nonumber \\
&+&\frac{d}{dt}\int_{0}^{l}\frac{1}{2}\kappa\left(x\right)\left( \frac{\partial u\left( x,t\right)}{\partial x}%
-1\right) ^{2}
\end{eqnarray}
Taking this last equation into equation (\ref{energia1}) it follows
after one integration more that
\begin{equation}\label{totalenergy}
E=T_{1}+T_{2}+\int_{0}^{l}\frac{1}{2}\rho \left( x\right) \left( \frac{%
\partial u\left( x,t\right)}{\partial t}\right) ^{2}dx+\int_{0}^{l}\frac{1}{2}\kappa\left(x\right)\left( \frac{%
\partial u\left( x,t\right)}{\partial x}-1\right) ^{2}dx
\end{equation}
This equation expresses the conservation of the total mechanical
energy of the system. The first three terms on the rhs of equation
(\ref{totalenergy}) represent the kinetic energy of the blocks and
of the massive spring, the last term represents the potential energy of
the spring. We can rewrite equation (\ref{totalenergy}) in terms of the $\xi\left(x,t\right)$
\begin{equation}\label{totalenergyxi}
E=T_{1}+T_{2}+\int_{0}^{\ell}\frac{1}{2}\rho \left( x\right) \left( \frac{%
\partial \xi\left( x,t\right)}{\partial t}\right) ^{2}dx+\int_{0}^{\ell}\frac{1}{2}\kappa\left(x\right)\left( \frac{%
\partial \xi\left( x,t\right)}{\partial x}\right)^{2}dx
\end{equation}
which turns out to be more useful in some applications.

Proceeding as in the case of the total linear momentum we can write the total energy in terms of the general solution. The result is
\begin{eqnarray}
E_{\mbox{\tiny total}}&=&\frac{1}{2}\left(M_1+M_2+M_{\mbox{\tiny total}}\right)V^2+\sum_{n=1}^\infty\left[M_1X_n\left(0\right)+M_2X_n\left(\ell\right)+\int_0^\ell\,\rho\left(x\right)X_n\left(x\right)\,dx\right]\dot T_n\left(t\right)\nonumber \\
&+&\sum_{n,m=1}^\infty\left[M_1X_n\left(0\right)X_m\left(0\right)+M_2X_n\left(\ell\right)X_m\left(\ell\right) +\int_0^\ell\,\rho\left(x\right)X_n\left(x\right)X_m\left(x\right) \,dx\right]\dot T_n\left(t\right)\dot T_m\left(t\right) \nonumber \\
&+&\int_0^\infty\,\kappa\left(x\right)\sum_{n,m=1}^\infty X_n^{\,\prime}\left(x\right)X_m^{\,\prime}\left(x\right) T_n\left(t\right)T_m\left(t\right)\,dx 
\end{eqnarray}
The last term representing the potential energy of the spring can be suitable rewritten with help of the following identity
\begin{equation}
\int_0^\ell\,\kappa\left(x\right)X_n^{\,\prime}\left(x\right)X_m^{\,\prime}\left(x\right)\,dx=\lambda_n\left[ M_1X_n\left(0\right)X_m\left(0\right)+M_2X_n\left(\ell\right)X_m\left(\ell\right)+\int_0^\ell\,\rho\left(x\right)X_n\left(x\right)X_m\left(x\right)\,dx \right]
\end{equation}
which can be easily proven. The final result is
\begin{eqnarray}
E_{\mbox{\tiny total}}&=&\frac{1}{2}\left(M_1+M_2+M_{\mbox{\tiny total}}\right)V^2+\sum_{n=1}^\infty\left[M_1X_n\left(0\right)+M_2X_n\left(\ell\right)+\int_0^\ell\,\rho\left(x\right)X_n\left(x\right)\,dx\right]\dot T_n\left(t\right)\nonumber \\
&+&\sum_{n,m=1}^\infty\left[M_1X_n\left(0\right)X_m\left(0\right)+M_2X_n\left(\ell\right)X_m\left(\ell\right) +\int_0^\ell\,\rho\left(x\right)X_n\left(x\right)X_m\left(x\right) \,dx \right]\dot T_n\left(t\right)\dot T_m\left(t\right) \nonumber \\
&+& \int_0^\infty\,\kappa\left(x\right) \sum_{n,m=1}^\infty \langle X_n\left(x\right)X_m\left(x\right) \rangle\,T_n\left(t\right) T_m\left(t\right)\,dx
\end{eqnarray}
Taking into account the orthonormality relation we can recast the total energy in the more illuminating form
\begin{equation}
E_{\mbox{\tiny total}}=\frac{1}{2}\left(M_1+m_2+M_{\mbox{\tiny spring}}\right)V^2+\sum_{n=1}^\infty\,\left[\dot T_n^2\left(t\right)+\omega_n^2\,T_n^2\right]
\end{equation}
or
\begin{equation}
E_{\mbox{\tiny total}}=\frac{1}{2}\left(M_1+m_2+M_{\mbox{\tiny spring}}\right)V^2+\sum_{n=1}^\infty\,\omega_n^2\left(\alpha_n^2+\beta_n^2\right)
\end{equation}
This last result shows that the total energy of the system can be decomposed into a sum of energies each one associated with a normal mode. Moreover, we can see that in order to excite two or more frequencies of comparable amplitudes it is necessary to supply the mode with the highest frequency with a greater amount of external energy. 
%

%
%
%
\section{Solution for $\rho\left(x\right) =0$}
When the spring is massless the motion of the two
blocks is easily obtained by reducing the two-body problem to the
the motion of a single effective body about a centre of force
\cite{Marion&Thornton1995}. Here we try to obtain those solutions by
making use of equations (\ref{eqt}) and (\ref{eqx}).

Firstly, notice that taking $\rho \left( x\right) =0,$ does not
eliminate the possibility of having eigenvalues different from zero.
It only means that the spatial eigenvalue equation is
\begin{equation}
\frac{d}{dx}\left[\kappa\left(x\right)\frac{dX\left(x\right)}{dx}\right]
=0
\end{equation}
the solution of which is
\begin{equation}\label{eqmrhozero}
X\left( x\right) =C\int_0^{x}\frac{dx^{\,\prime}}{\kappa\left( x^{\,\prime}\right)
}+X\left( 0\right)
\end{equation}
The position of a point of the spring is then given by
\begin{equation}
u\left( x,t\right)
= x_{0}+ x + Vt + \left[C\int_0^{x}\frac{dx^{\,\prime}}{\kappa\left( x^{\,\prime}\right)
}+X\left( 0\right) \right]\,A\cos\,(\omega t + \phi) 
\end{equation}
Defining the usual spring constant $K_e$ by
\begin{equation}
K_{e}^{-1}=\int_{0}^{\ell}\frac{dx}{\kappa\left(x\right) }
\end{equation}
with the help of equation (\ref{eqmrhozero}) we obtain
\begin{equation}
C=K_{e}\left[ X\left(\ell\right) -X\left( 0\right) \right]
\end{equation}
Making use of the boundary conditions in the form given by equations
(\ref{cc1}) and (\ref{cc2}) we have
\begin{eqnarray}
K_{e}X\left(\ell\right) -\left( K_{e}-\omega ^{2}M_{1}\right)
X\left( 0\right)
&=&0 \\
\left( K_{e}-\omega ^{2}M_{2}\right) X\left( \ell\right)
-K_{e}X\left( 0\right) &=&0
\end{eqnarray}
In order to have a non-trivial solution the determinant associated
with this linear system must be zero, that is
\begin{eqnarray}
M_{1}M_{2}\,\omega ^{4}-K_{e}\left(M_1+M_2\right)\,\omega ^{2} &=& 0
\end{eqnarray}
It follows that the allowed eigenfrequency is as expected given by
\begin{equation}
\omega_1 =\sqrt{\frac{K_{e}}{\mu }}
\end{equation}
where
\begin{equation}
\mu =\frac{M_{1}M_{2}}{M_{2}+M_{1}}
\end{equation}
is the reduced mass of the system block 1 plus block 2. This
eigenfrequency and the zero mode frequency $\omega_0=0$ are the only
allowed frequencies of the system when the spring is massless.
Making use of equation of our definition of the scalar product we can calculate easily the constant $C$ that normalises the eigenfunction. The result is
\begin{equation}
C=\frac{K_e}{\sqrt{\mu}}
\end{equation}
%
%
%
%
%
\section{Solution for $\kappa\left(x\right)$ and $\rho\left(x\right)$ uniform}
We now turn our attention to an important special case. When the
elastic function of the spring and its density are uniform it is possible to solve analytically the equation of
motion, i.e. the wave equation that describe the system and 
interpret clearly the solutions. Defining
\begin{equation}
v^{2}:=\kappa/\rho
\end{equation}
and with $\lambda:=\omega^2$ to be in accordance with the standard
notation, equations (\ref{eqt}) and (\ref{eqx}) read
\begin{eqnarray}
\frac{d^{2}T}{dx^{2}}+\omega ^{2}T &=&0 \\
\frac{d^{2}X}{dx^{2}}+q^{2}X &=&0
\end{eqnarray}
where
\begin{equation}
\omega =qv.  \label{wqv}
\end{equation}
The boundary conditions, equations (\ref{cc1ksi}) and (\ref{cc2ksi}) or (\ref{cc1}) and (\ref{cc2}) applied to this particular case lead to
\begin{eqnarray}
M_{1}X\left( 0\right) q^{2}v^{2}+\kappa X^{\prime }\left( 0\right) &=&0 \\
M_{2}X\left(\ell\right) q^{2}v^{2}-\kappa X^{\prime
}\left(\ell\right) &=& 0
\end{eqnarray}
The general solution for the spatial part is
\begin{equation}
X\left( x\right) =A\cos qx+B\sin qx
\end{equation}
The allowed eigenvalues are determined by the linear algebraic
system
\begin{eqnarray}
M_{1}qv^{2}A+kB &=&0 \\
\left( M_{2}qv^{2}\cos q\ell+\kappa\sin q\ell\right) A+\left(
M_{2}qv^{2}\sin q\ell-\kappa\cos q\ell\right) B &=&0
\end{eqnarray}
whose characteristic equation is
\begin{equation}
\tan q\ell=\frac{\rho q}{\mu q^{2}-\displaystyle{\frac{\rho ^{2}}{M_{1}+M_{2}}}}
\label{caracteristica1}
\end{equation}
To illustrate the discussion let us consider the situation for which
the density of the spring is very small. In this case the mass of
the spring can be neglected. Making the necessary approximations to
equation (\ref{caracteristica1}) we obtain
\begin{equation}
q\approx\sqrt{\frac{\rho }{\mu\ell}} \label{q1}
\end{equation}
The angular frequency  is given by equation (\ref{wqv}) and in this
case it leads to
\begin{eqnarray}
\omega\approx \sqrt{\frac{k}{\mu \ell}}
\end{eqnarray}
As expected the last result has a non-trivial limit as the mass density tends to zero. Notice that $\kappa/\ell$ can be identified with the usual elastic 
constant of the spring, Notice also that the speed of the wave does depend on the
density of the spring and tends to infinity as the mass density
tends to zero. It is precisely this fact that in this approximation
makes possible to replace the real forces by forces between the two point masses obeying Newton's action and reaction principle discussed in the introduction. To investigate the next order correction to the angular frequency we consider for simplicity the case where one of the point masses, say $M_1$, is infinite and the total mass of the point particles $M_1+M_2$ is also infinite. This situation corresponds to the case where one of the extremities of the spring is fixed to a wall. Adding one more term to the
expansion of $\tan q\ell$ in (\ref{caracteristica1}) we obtain the following quartic equation for $q$
\begin{equation}
\frac{1}{3}q^4+\frac{1}{\ell^2}q^2-\frac{\rho}{M_2\ell^3}=0
\end{equation}
whose physical solution is given by
\begin{equation}
q=\sqrt{\frac{\rho }{M_2\ell}}-\frac{1}{6}\sqrt{\frac{\ell\rho
^{3}}{M_2^{3}}}
\end{equation}
Consequently, we will have
\begin{equation}
\omega =\left( \sqrt{\frac{\rho}{M_2
\ell}}-\frac{1}{6}\sqrt{\frac{\ell\rho^{3}}{M_2^{3}}}\right)
\sqrt{\frac{\kappa}{\rho}}
\end{equation}
A little bit more of simple algebra allows to write
\begin{equation}
\omega\approx \sqrt{\frac{\kappa}{\ell\left(M_2
+\frac{1}{3}m\right)}}
\end{equation}
a well known result, see for example \cite{French1971}, \cite{GalloniKohen}.
\begin{figure}[!t]
\begin{center}
\includegraphics[height=10cm, width=10cm]{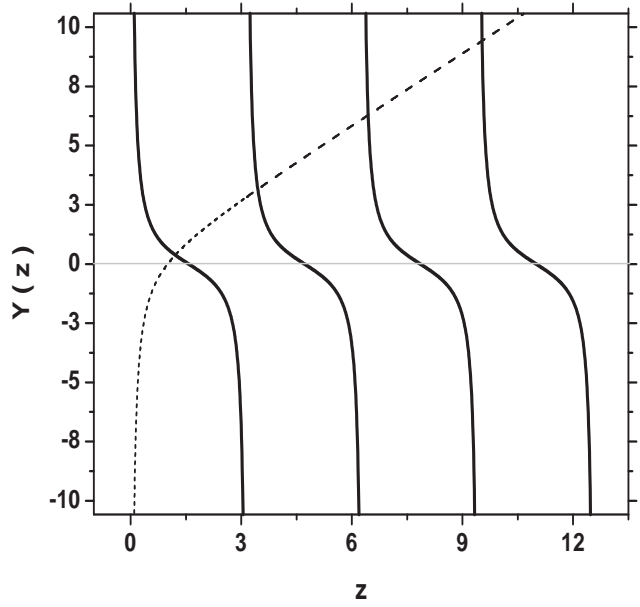}
\caption{The dashed curve is the graphical representation of the rhs of equation (\ref{char}) for given $m$, $M_1$ and $M_2$. }
\label{cotangent}
\end{center}
\end{figure}
\section{The angular eigenfrequencies}
In order to investigate a general solution of equation ( \ref{caracteristica1}) let us define the
variable $z=q\ell$ and write the characteristic equation (\ref{caracteristica1}) in the form
\begin{equation}\label{char}
\cot z=\frac{\mu z}{m}-\frac{m}{Mz}
\end{equation}
%
\begin{figure}[!t]
\begin{center}
\includegraphics[height=10cm, width=10cm]{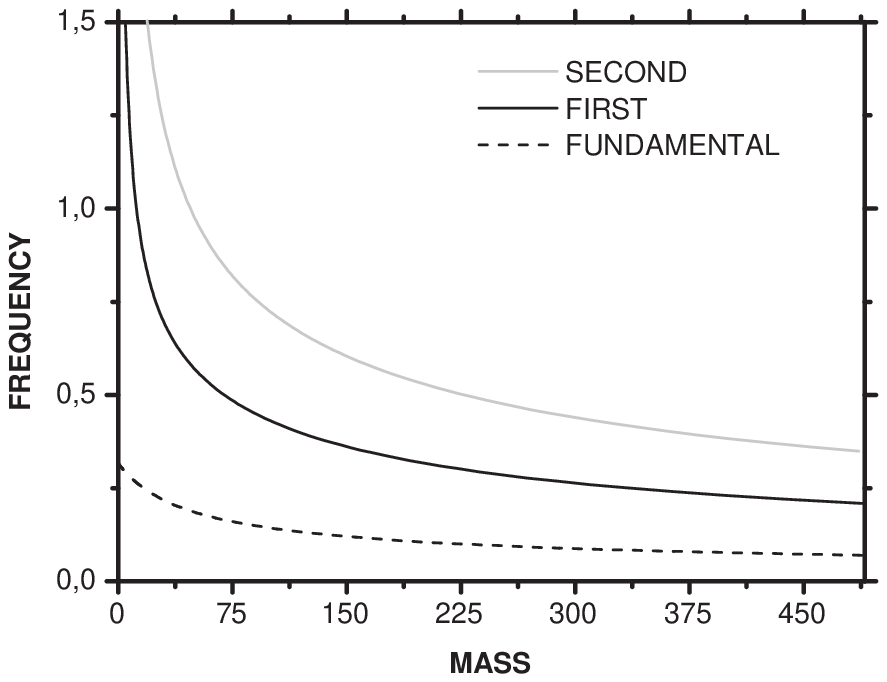}
\caption{The curves show the behaviour of the frequency of the first three modes of the system as a function of the mass of the spring. The sum of the attached masses is same for the three curves and in this case one of the attached masses is a fixed wall. }
\label{graphfive}
\end{center}
\end{figure}
\begin{figure}[!t]
\begin{center}
\includegraphics[height=10cm, width=10cm]{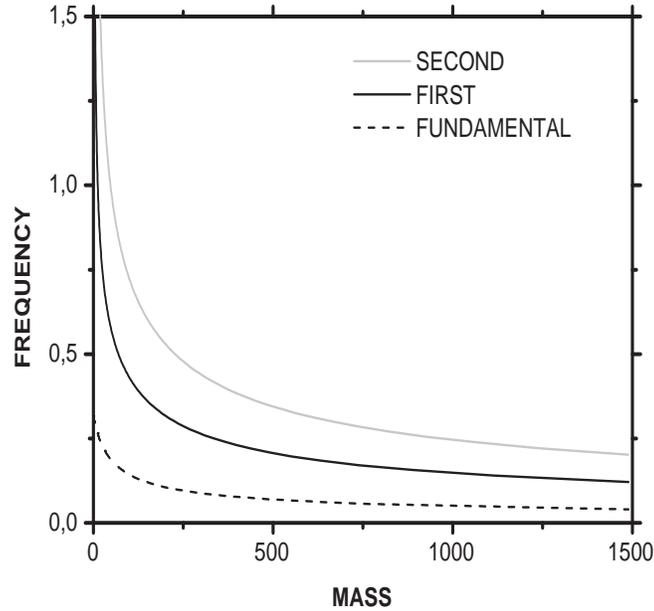}
\caption{The curves show the behaviour of the frequency of the first three modes of the system as a function of the mass of the spring. One of the attached masses is a fixed wall. }
\label{graphsix}
\end{center}
\end{figure}

\begin{figure}[!t]
\begin{center}
\includegraphics[height=10cm, width=10cm]{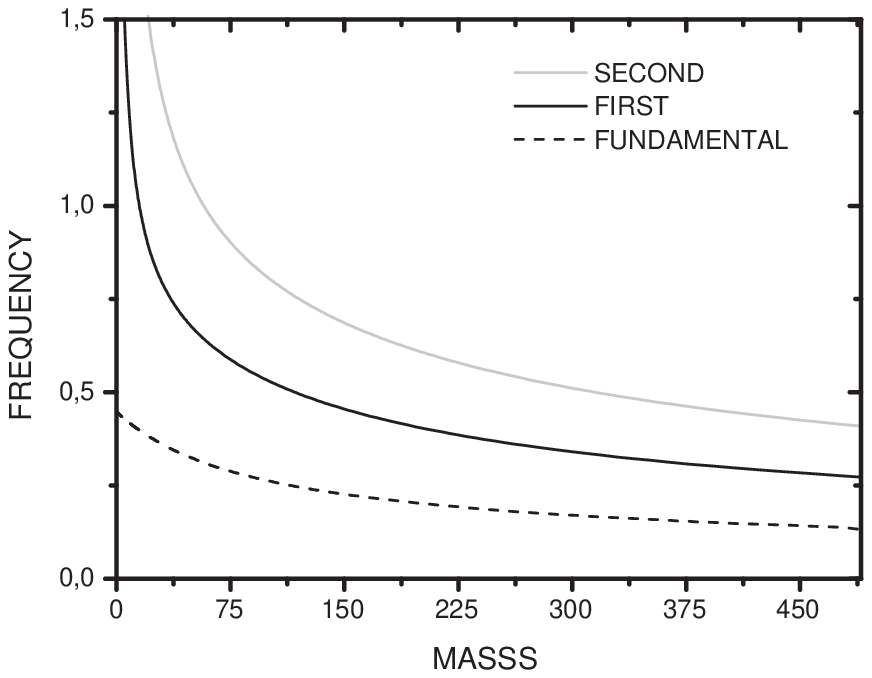}
\caption{The curves show the behaviour of the frequency of the first three modes of the system as a function of the mass of the spring. The sum of the attached masses is same for the three curves and in this case both attached masses are finite. }
\label{graphseven}
\end{center}
\end{figure}

where $M:=M_1+M_2$. In Figure \ref{cotangent} we plot the lhs and the rhs of equation (\ref{char}) separately for representative values of $\mu, M$ and $m$. The solutions of the charcteristic equation are determined by the intersection points. It is easily seen that there is an infinite number of eigenfrequencies, one in each open interval $(n\pi, (n+1)\pi)$, where $n$ is a non-negative integer. The lowest eigenfrequency lies in the interval $(0,\pi)$. The lowest eigenfrequency is the only one that remains finite when the mass of the spring tends to zero. All other eigenfrequencies tend to infinite and this means that they are increasingly harder to excite. For $n \gg 0$ the highest eigenfrequencies can be approximately described by the simple formula $z_n=n\pi$. Then we can write
\begin{equation}
\omega_n\approx n\pi\sqrt{\frac{\kappa}{\rho\ell^2}}=n\pi\sqrt{\frac{K_e}{m} }
\end{equation}
In order to obtain an analytical approximate solution for the eigenfrequencies we solve  equation (\ref{char}) for $m$ to obtain
\begin{equation}
m=\frac{1}{2\tan z}\left( -M \pm \sqrt{\left(
M^{2}+4\left( \tan ^{2}z\right) \mu M\right) }\right) z.
\end{equation}
where the plus sign must be used if $x\in (n\pi, n\pi+\pi/2)$ and the minus sign when if $z\in (n\pi+\pi/2), (n+1)\pi$.
Now we define $w=\sqrt{m}$ and make use of the B\"{u}rmann- Lagrange theorem \cite{Tikhonov} to express the inverse function in the series form. The result is
\begin{equation}
z =\frac{1}{\sqrt{\mu }}\left[ w+\left( -\frac{1}{6\mu }+\frac{1}{2M}%
\right) w^{3}+\left( \frac{11}{360\mu ^{2}}-\frac{1}{12\mu M}-\frac{1}{8M^{2}%
}\right) w^{5}+\cdots\right]
\end{equation}
Consider only the first term of this series. Then it is easily seen
that
\begin{equation}
q \approx \frac{1}{\ell}\sqrt{\frac{m}{\mu }}
\end{equation}
In this case the angular frequency is
\begin{equation}
\omega_0\approx\sqrt{\frac{\kappa}{\mu }}
\end{equation}
Let us consider the first correction to this result which means to
take into account the term in $w^3$ in the inverted series. Then it
follows that
\begin{equation}
q \approx \frac{1}{\ell}\sqrt{\frac{m}{\mu}}\left[ 1+\left( -\frac{1}{6\mu }+\frac{1}{2M}\right)
m\right]
\end{equation}
The angular frequency is then
\begin{equation}
\omega_0\approx \sqrt{\frac{\kappa}{\mu}}\left[\left( 1-\frac{1}{6\mu }+\frac{1}{2M}%
\right) m\right]
\end{equation}
We can also express the other eigenfrequencies ($n>0$) in a series form by using again the B\"urmann-Lagrange theorem. The result up to the fourth power in the mass of the spring is
\begin{eqnarray}
\omega_n &=& n \pi\sqrt{\frac{K_e}{m\ell}}\left[1+\frac{m }{\mu n^2\pi^2}-\frac{m^2}{\mu^2 n^3\pi^3}+\left(\frac{2}{\mu^3 n^6\pi^6} -\frac{1}{3\mu^3 n^4\pi^4}
+\frac{1}{\mu^2 n^4\pi^4 M}\right)m^3\right.\nonumber \\
&+&\left.\left( -\frac{5}{\mu^4 n^8\pi^8}+\frac{4}{3\mu^4 n^6\pi^6 M}-\frac{4}{\mu^3 n^6\pi^6 M}\right)m^4\right]
\end{eqnarray}
Figures \ref{graphfive}, \ref{graphsix}, and \ref{graphseven}} show the behaviour of the first three lowest eigenfrequencies as a function of the mass of the spring $m$ for a particular choice of the sum of the attached masses  $M_1+M_2$. In Figures \ref{graphfive} and \ref{graphsix} one of the attached masses is infinite and the other one is finite. This means that one of the ends of the spring is attached to a fixed wall.  In Figure \ref{graphseven} both attached masses are finite. As physically expected, when the mass of the spring goes to zero the higher modes become harder and harder to excite and the fundamental mode tends to a fixed value. 
%
\section{Final remarks}
In this paper we discussed the classical mechanics of a spring of
arbitrary mass coupled to two arbitrary massive blocks attached at its ends. A
general approach to the problem was attempted and some general results such as the conservation of linear momentum and energy
were obtained. We have shown also that the physical problem leads to an example of a Sturm-Liouville system. The detailed study of this problem is heavily dependent on the explicit knowledge of the elastic function
$\kappa\left(x\right)$. The special case for which the elastic
function and the mass density are uniform was discussed and an approximation procedure to the evaluation of the normal frequencies was put forward and tested. In the limiting case of a massless spring, we have focused our attention on the motion of the attached masses $M_1$ and $M_2$, and considered the spring as a way of transmitting the interection between them. With the respect to the wave motion of the spirng, we observe that the result $\sqrt{K_e/\rho}$ is the velocity of the wave only if the velocity of the centre of mass of the system is zero. If this velocity is $V$ with respect to some suitable reference frame then accordding to the galilean rule the velocity of pulse propagation will be $V=\sqrt{K_e/\rho}$.

At the moment, the study of a possible equivalence between motion in a single mode of the massive spring and simple harmonic motion and possible quantisation of the system is under way.
\section*{Acknowledgments}
Two of us  (Y A C and L R-P) wish to acknowledge the financial help of FAPERJ, Funda\c c\~ao de Amparo \`a Ci\^encia do Estado do Rio de Janeiro.
%

%

\begin{thebibliography}{99}
%
\bibitem{French1971}French A P 1971 \emph{Vibrations and Waves}, (New York: Norton)

\bibitem{Marion&Thornton1995}  Marion J B and Thornton S T 1995 \emph{Classical Dynamics of Particles and Systems},
5th edn, (Orlando: Saunders College Publishing)

\bibitem{GalloniKohen} Galloni E E  and Kohen M 1979 Am. J. Phys. \textbf{47} 1076

\bibitem{Landau}  Landau L D and Lifshits E 1995 \emph{Th\'eorie de L'Elasticit\'e}, (Moscow: Mir)

\bibitem{Tikhonov} Tikhonov A N and Samarskii A A 1996 \emph{ Equations of Mathematical Physics}, (New York: Dover)
\end{thebibliography}
\end{document}